\documentclass[twocolumn,twoside,slac_two]{revtex4}
\usepackage{graphicx}
\usepackage{fancyhdr}
\def\posttrials{7.5 standard deviations}

\def\specradius{$0.24^{\circ}$}

\def\fittedcentroidsexg{R.A. $20^{h} 20^{m} 04.8^{s}$, Decl. $\mathrm +40^{\circ} 45^{\prime} 36^{\prime\prime}$ (J2000)}
\def\fittedextension{$\rm 0.23^{\circ} \pm {0.03^{\circ}_{\mathrm stat}} {}^{+0.04^{\circ}}_{-0.02^{\circ}{\mathrm sys}}$}

\def\threshold{320~GeV}

\def\specindexg{$\rm \Gamma = 2.37 \pm {0.14}_{\mathrm stat} \pm {0.20}_{\mathrm sys}$}

\def\fluxnorm{$\rm N_{0} = (1.5 \pm 0.2_{stat} \pm {0.4}_{sys} ) \times 10^{-12} \ ph  \ {TeV}^{-1} \ cm^{-2} \ s^{-1}$}
\def\integralflux{$\rm 5.2 \pm 0.8 _{stat} \pm {1.4}_{sys} \times 10^{-12} \ ph \ cm^{-2} \ s^{-1}$}

\def\fluxpercentage{3.7\%} 

\def\inferredensities{$1.0-5.5$~cm$^{-3}$}

\def\centroidpulsaroffset{$\sim 0.5^{\circ}$}

\def\snr{SNR~G78.2+2.1}
\def\ver{VER~J2019+407}

\def\fgl{1FGL~J2020.0+4049}
\def\fgltwo{2FGL~J2019.1+4040}
\def\fglpsr{1FGL~J2021.5+4026}
\def\psr{PSR~J2021+4026}

\newcommand\veritas{{VERITAS}}

\newcommand\fermi{{\it Fermi}}

\newcommand{\mnras}{MNRAS}
\newcommand{\aj}{AJ}
\newcommand{\apj}{ApJ}
\newcommand{\apjs}{ApJS}

\newcommand{\aap}{AAP}
\newcommand{\aaps}{AAPS}

\begin{document}

\title{VERITAS Observations of the Vicinity of the Cygnus Cocoon}

\author{A. Weinstein for the VERITAS Collaboration}
\affiliation{Iowa State University, 1 Osborne Dr., Ames, IA, 50011, USA}

\begin{abstract}

The study of $\gamma$-ray emission from galactic sources such as supernova remnants
(SNR) may provide key insights into their potential role as accelerators of cosmic rays up to
the knee ( $\sim 10^{15}$ eV).  The VERITAS Observatory is sensitive to galactic and extragalactic
$\gamma$-ray sources in the 100 GeV to 30 TeV energy range.  We report here on VERITAS
observations of the vicinity of the cocoon of freshly accelerated cosmic rays reported by
Fermi, which lies between potential accelerators in the Cygnus OB2 association and the
$\gamma$-Cygni SNR.  A particular focus is placed on the source VER J2019 +407 in $\gamma$-Cygni.

\end{abstract}

\maketitle

\thispagestyle{fancy}


\section{The VERITAS Instrument}

VERITAS, located at the Fred Lawrence Whipple Observatory near Tucson, Arizona, is an array of four 12-meter imaging atmospheric Cherenkov telescopes.  Each telescope has a pixelated camera comprised of 499 photomultiplier
tubes with a $3.5^{\circ}$ field of view.  Designed to detect photons of astrophysical origin between 100 GeV and 30 TeV, VERITAS detects and images the secondary Cherenkov light produced when gamma rays and cosmic rays initiate particle cascades in the upper atmosphere.  Stereoscopic reconstruction of events using multiple telescopes allows for single-photon angular resolution of better than $0.1^{\circ}$ and energy resolution on the order of 15-25\%. In its current configuration, VERITAS can detect a
source with $\sim 1\%$ of the Crab Nebula flux in less than 30 hours.
The instrument has been operating in full array mode since 2007.  See \citep{2006APh....25..391H} for  further details regarding the operation of VERITAS.

\section{The $\gamma$-Cygni Supernova Remnant}

SNR~G78.2+2.1, also known as the $\gamma$-Cygni supernova remnant (SNR), is a $\sim\!1^{\circ}$ diameter
shell-like radio and X-ray SNR\citep{Higgs1977,Lozinskaya2000}.  It is estimated to be at a distance of $\sim 1.7$~kpc\citep{Higgs1977,Lozinskaya2000} and to be approximately $5000-7000$ years old\citep{Higgs1977,Landecker1980}.
It is thought to be in an early phase of adiabatic expansion into a low-density medium\citep{Lozinskaya2000}.
A slowly expanding {\sc H i} shell immediately surrounding the radio shell was found by \citet{Gosachinskij2001}, which
\citet{Lozinskaya2000} believes to have been created by the progenitor stellar
wind.

The radio shell divides roughly into northern and southern arcs \citep{Zhang1997,uchiyama}.
Enhanced thermal X-ray emission in the north suggests shocked gas \citep{uchiyama} and falls in
a void of CO emission \citep{landp}.
Strong optical emission with sulfur lines also characterizes the region \citep{Mavromatakis2003}.
The $\gamma$-ray satellite \fermi\/, operating at GeV energies, has discovered a $\gamma$-ray pulsar \psr\ at the center of the remnant \citep{onefgl,fermipulsarcatalog}.  While this pulsar has a low
luminosity ($1.1\times 10^{35}$ ~erg~s$^{-1}$) and a spin-down age (76.8~kyr)
much greater than the estimated age of \snr\/, the pulsar kinematics make it
probable that \psr\ was born with something close to its current spin period
and is the remnant of {\snr}'s progenitor star \citep{Trepl2010}.  Diffuse $\gamma$-ray emission above 10 GeV is also reported by \fermi\ over the full extent of the remnant \citep{Lande}.  A point source co-located with \ver\ was previously reported in the first and second \fermi\ catalogs \citep{onefgl,twofgl}, but \citet{Lande} conclude it to be artifact.

\section{The Cygnus Cocoon}

\citet{2011Sci...334.1103A} reported an extended region of emission above a
few GeV that they interpreted as a freshly accelerated cocoon of cosmic rays. The
cavity defining the cocoon region is outlined by ionization fronts visible in the mid-
infrared.  $\gamma$-Cygni overlaps one end of the cocoon and is a potential source
of the cocoon's trapped cosmic rays.

\begin{figure*}[t]
\includegraphics[width=135mm]{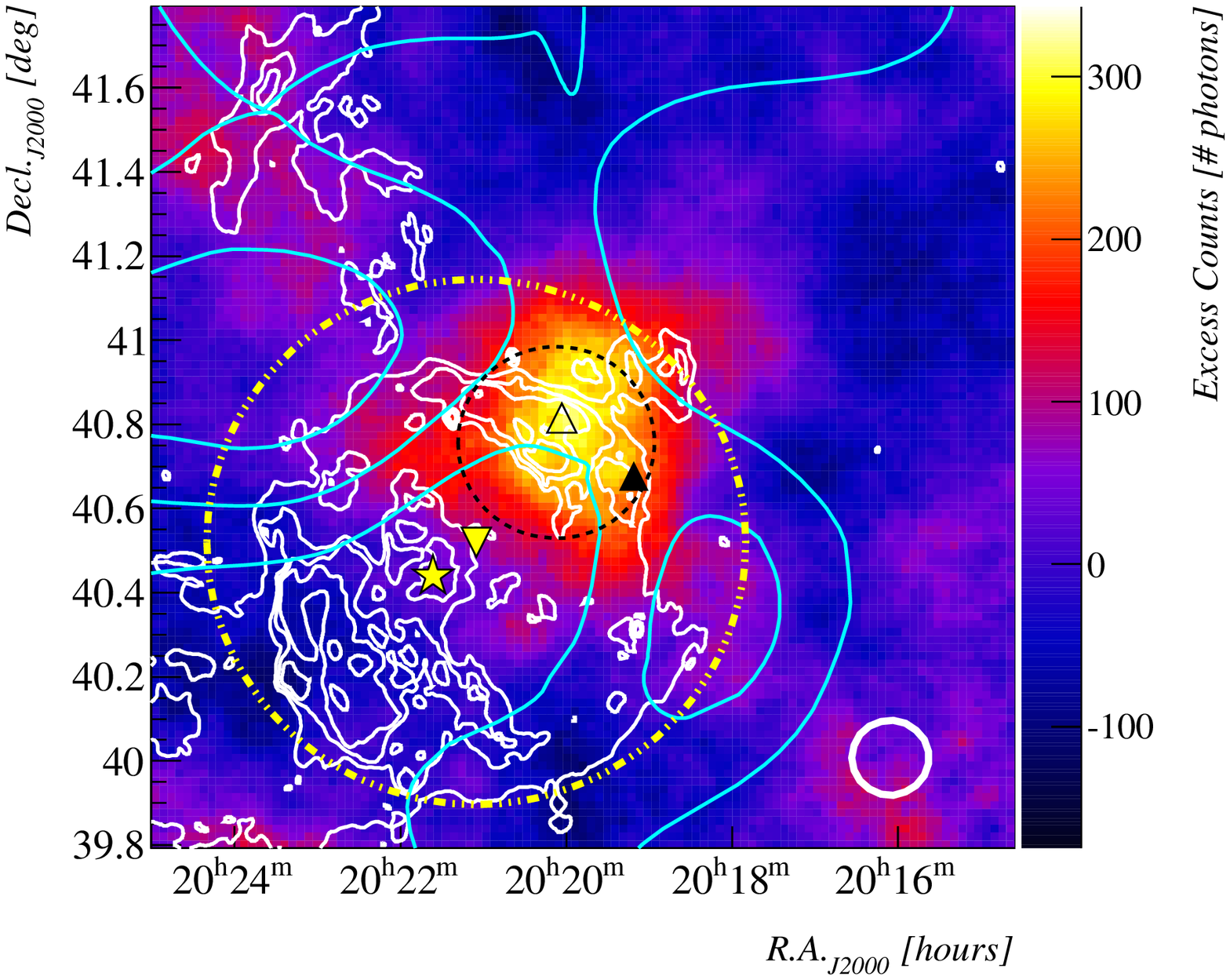}
\caption{\veritas\ $\gamma$-ray image of \snr\ showing the detection of \ver\ and its fitted extent (black dashed circle).    CGPS 1420 MHz continuum radio contours at brightness temperatures of 23.6K, 33.0K, 39.6K, 50K and 100K (white)
\citep{CGPS} outline the radio SNR; the star symbol shows the location of the
central $\gamma$-ray pulsar \psr.  The inverted triangle and dot-dashed circle (yellow) show the fitted centroid and extent of the emission detected by \fermi\ above 10 GeV.
The open and filled triangles (black) show the positions of \fermi\ catalog sources \fgl\ and \fgltwo\/.  The 0.16, 0.24, and 0.32 photons/bin contours of the \fermi\ detection of the Cygnus cocoon are shown in cyan.
The white circle (bottom right
corner) indicates the 68\% containment size of the \veritas\ $\gamma$-ray PSF for this analysis.
\label{fig:excesswithoverlay} }
\end{figure*}

\section{Detection of $\gamma$-ray Emission from the Direction of $\gamma$-Cygni}

Figure~\ref{fig:excesswithoverlay} displays the acceptance-corrected very-high-energy (VHE) $\gamma$-ray excess map of the area around \snr\/ \citep{VERJ2019}.  An extended source is seen overlapping the northern edge of the remnant, with a detection significance of \posttrials\/.
We use a binned, extended maximum-likelihood fit to the raw counts map to assess the \ver\ morphology.  The source is modeled in the fit as a symmetric, two-dimensional Gaussian convolved with the \ver\ point-spread function (PSF); the background is assumed to be flat before exposure effects are taken into account.
We find a fitted extension of the two-dimensional Gaussian to be \fittedextension\/, with fitted centroid coordinates \fittedcentroidsexg\/.
The statistical uncertainty in the centroid location is $0.03^{\circ}$, with a systematic uncertainty of $0.018^{\circ}$.  The systematic uncertainty considers both the telescope pointing error and systematic errors of the fit itself \citep{VERJ2019}.
The positions of the $\gamma$-ray pulsar \psr\ (\fglpsr) (\centroidpulsaroffset\ from \ver\ ) and the centroid of the emission above 10 GeV from the remnant (as seen by \fermi\ ) are also shown for reference.

\begin{figure}
\includegraphics[width=80mm]{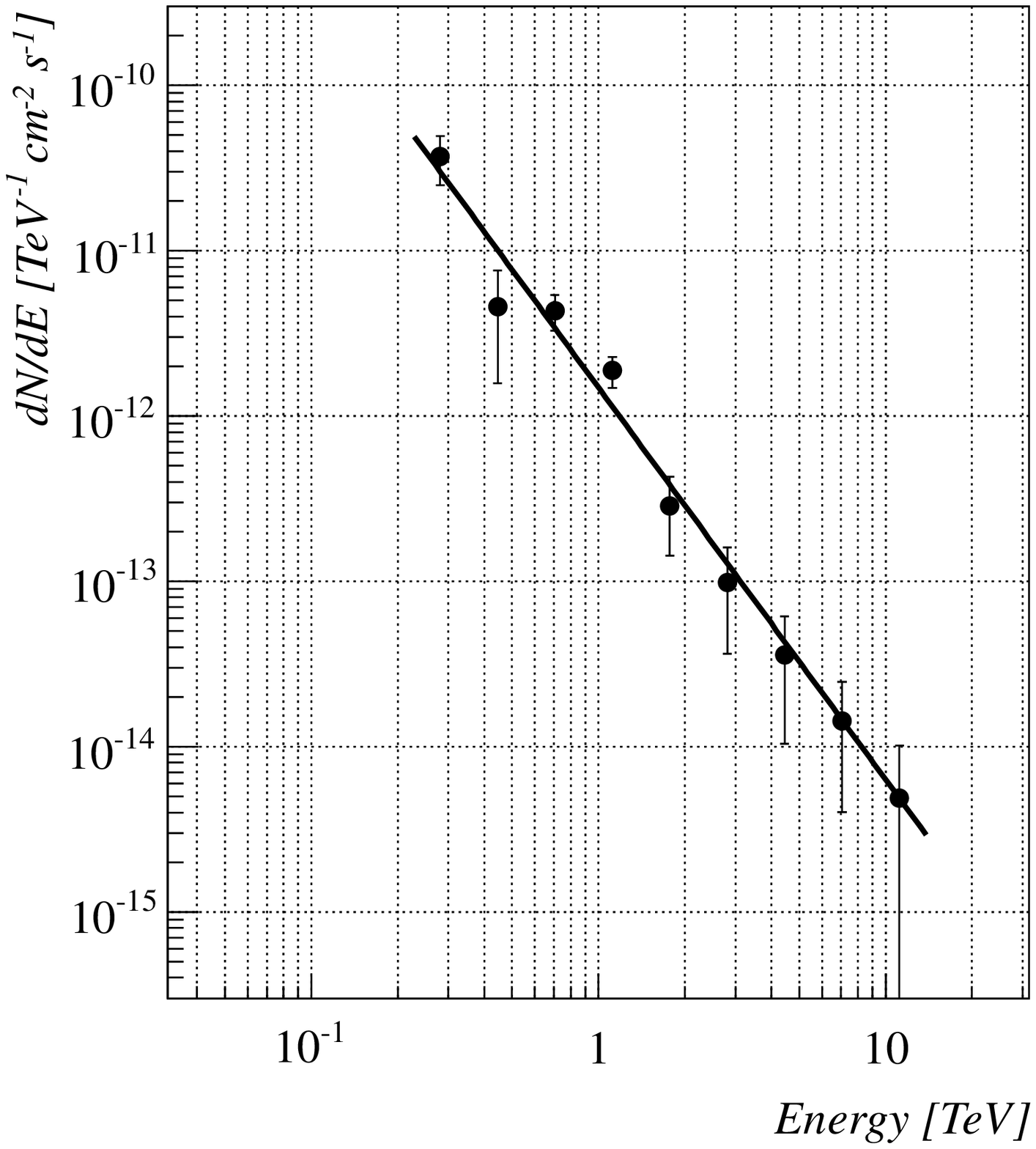}
\caption{Spectrum of
\ver, derived from 4-telescope data only.  Points are the \veritas\ spectrum; the line is a power-law fit with a spectral index of \specindexg\ and a flux
normalization of \fluxnorm\/.\label{fig:tevspec}}
\end{figure}

Figure \ref{fig:tevspec} shows the spectrum of reconstructed $\gamma$-ray events within \specradius\/ from R.A. $\rm 20^{h}19^{m}48^{s}$, Decl. $\rm +40^{\circ}54'00''$.
Runs where only three of four telescopes were operational have been excluded from this sample \citep{VERJ2019}.  The photon spectrum is consistent with a differential power law in energy, $\rm dN/dE = N_{0}\times (E/TeV)^{-\Gamma}$, between the analysis threshold of  \threshold\/ and 10 TeV.
The photon index is \specindexg\/ and the flux normalization at 1 TeV is {\fluxnorm}.  The integral flux above \threshold\/ (\integralflux) corresponds to \fluxpercentage\/ of the Crab Nebula flux above that energy \citep{VERJ2019}.

\subsection{$\gamma$-Cygni in X-Rays}

\begin{figure}[t]
\includegraphics[width=80mm]{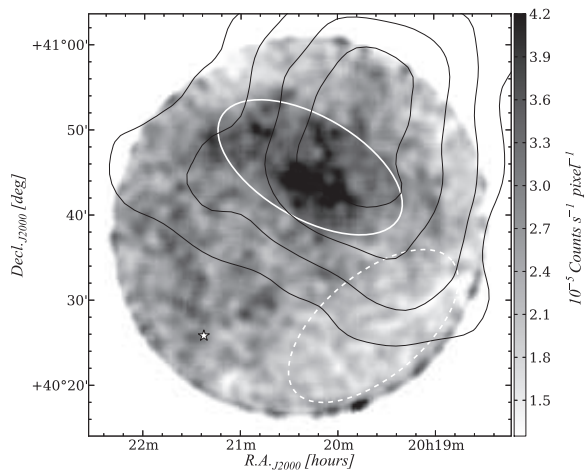}
\caption{ ASCA X-ray view of G78.2+2.1 between 1 and 3 keV, overlaid with the \ver\ smoothed photon excess contours (100, 150, 210 and 260 photons).  The region used to extract a spectrum and the corresponding background region to the south of the remnant are indicated by white solid and dashed ellipses, respectively.  A white star marks the position of \psr .}
\label{fig:asca}
\end{figure}

Figure~\ref{fig:asca} illustrates the region of enhanced X-ray emission overlapping \ver\/ \citep{VERJ2019}.  It displays the $0.7-3.0$ keV exposure-corrected X-ray map using data from ASCA Sequence \#25010000 (data originally presented by \citet{uchiyama}), generated by co-adding data from the two gas imaging spectrometers.  A spectrum was extracted from a $12^{\prime}\times 24^{\prime}$ elliptical region enclosing most of the X-ray emission inside the \veritas\ contours, centered on coordinates R.A.~$20^{\rm h}$~$20^{\rm m}$~$17^{\rm s}$, Decl.~$+40^{\circ}$~$45^{\prime}$~$41^{\prime\prime}$ (J2000) and oriented with position angle $60^{\circ}$.  We selected background photons from an identically sized ellipse near the center of the remnant at R.A.~$20^{\rm h}$~$19^{\rm m}$~$38^{\rm s}$, Decl.~$+40^{\circ}$~$27^{\prime}$~$02^{\prime\prime}$ (J2000), with position angle is $130^{\circ}$. The source and background regions are displayed in Figure~\ref{fig:asca}.

\begin{figure}[t]
\includegraphics[width=80mm]{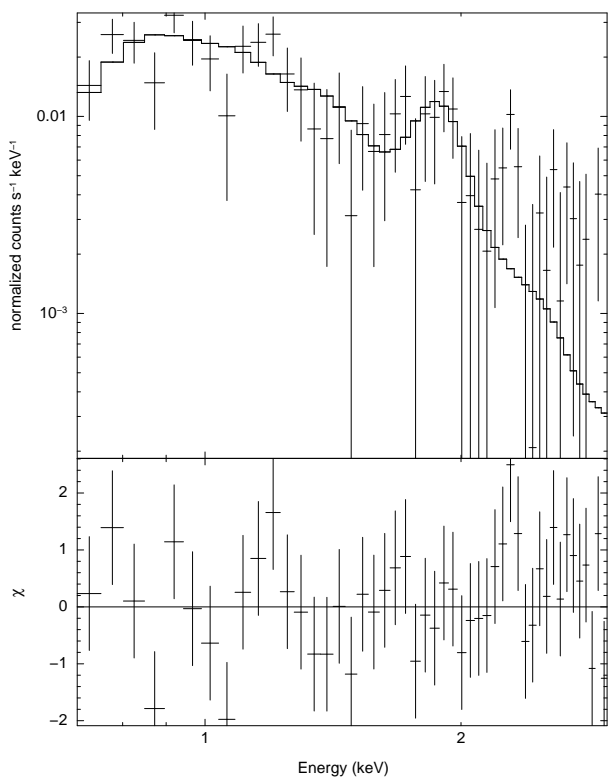} 
\caption{
Top Panel --- ASCA X-ray spectrum of the region of enhanced X-ray emission coincident with \ver\ shown in Figure~\ref{fig:asca}.
The stepped line shows the fit of a Raymond-Smith thermal plasma model with parameters as given in the text.
Bottom Panel --- residuals from the best-fit model.}
\label{fig:ascaspec}
\end{figure}

The source spectrum between $0.7$ and $3.0$ keV is shown in Figure~\ref{fig:ascaspec} \citep{VERJ2019}.  The X-ray spectrum was modeled using an absorbed Raymond-Smith thermal plasma model with a best-fit temperature of $kT = 0.57 \pm 0.14$ keV.  With this model the column density is $N_H = \rm (3.7 \pm 2.0) \times 10^{21} \thinspace cm^{-2}$, the normalization is $\rm N=1.8\times 10^{-3} cm^{-5}$m and the absorption-corrected flux is $6.0 \times 10^{-12} \rm \, erg \, cm^{-2} \, s^{-1}$ in the 0.5$-$8.0~keV band.  This result differs significantly from that given in \citep{uchiyama}, which claimed an additional power-law component and a large Ne IX line feature.
The divergence in results appears to hinge, not on the choice of source region, which is similar, but on the choice of background region \citep{VERJ2019}. We selected a background region that was as close as possible to the source while \citep{uchiyama} chose a region $3.5^{\circ}$ away.  We can produce Uchiyama's results by choosing a background region similar to Uchiyama's.  For further details, see \citet{VERJ2019}.

\section{Interpretation}

It is plausible that the VHE $\gamma$-ray emission seen from $\gamma$-Cygni arises from particles accelerated in shocks occuring at the interaction of the supernova ejecta and the surrounding medium.
These particles could be either accelerated electrons, which would produce the emission via inverse-Compton scattering, or accelerated nuclei.  Should they be high-energy electrons, they would also be expected to produce X-ray synchrotron radiation, which would appear as a non-thermal power-law component in the X-ray spectrum.  While our analysis of the ASCA X-ray spectrum does not argue for a non-thermal component, our upper limit on this component is still weak enough that we cannot exclude the possibility that the TeV, if not the GeV, emission is due to inverse-Compton scattering \citep{VERJ2019}.
On the other hand, it is also plausible that the VHE $\gamma$-ray emission is produced by interaction of accelerated nuclei with the {\sc H{\thinspace}i} shell surrounding the remnant.  Estimates of the target material density required for accelerated nuclei to produce the observed VHE $\gamma$-ray flux, based on \cite{drury1994}, give a range of densities \inferredensities\/ \citep{VERJ2019} consistent \citet{Gosachinskij2001}'s estimates of the gas density within the {\sc H{\thinspace}i} shell.
However, it must be noted that the shock velocities inferred from the optical and X-ray data are too low for the forward shock to be currently accelerating particles to TeV energies.  If the VHE $\gamma$-ray emission is hadronic, it is likely due to particles accelerated when the remnant was younger that are now interacting with the shell.

The relationship of the cocoon of freshly-accelerated cosmic rays detected by \fermi\ to \snr\ and \ver\ also remains unclear.  It is possible \snr\ either has injected or is injecting accelerated particles into the cocoon.  However, while they are shown for reference, we caution against using the cocoon contours from \citet{2011Sci...334.1103A} to judge the relationship of the cocoon to the VHE $\gamma$-ray emission, since they are derived from an analysis where \fgl\/, which is no longer considered an independent source, was included as part of the background model.
It should also be noted that the VHE $\gamma$-ray excess map in this paper was made with the ring-background estimation method, which is ill-suited to detecting a large-scale ($\sim 4$ square degree) region of $\gamma$-ray emission such as the cocoon.
Therefore the VHE $\gamma$-ray maps shown here cannot be used to set a meaningful upper limit on cocoon emission above 300 GeV.
A conclusion determination of the relationship between \snr\ and \ver\ will have to await further data, analyzed with more sophisticated analysis techniques.




\bigskip 
\begin{acknowledgments}

This research was supported by grants from the U.S. Department of Energy, the U.S.
National Science Foundation and the Smithsonian Institution, by NSERC in Canada, by
Science Foundation Ireland and by STFC in the UK . We acknowledge the work of the
technical support staff at the Fred Lawrence Whipple Observatory and at the
collaborating institutions in the construction and operation of the instrument. Amanda
Weinstein and Vikram Dwarkadas acknowledge the support of NASA grant NNX11AO86G.  The Cygnus cocoon
contours were graciously provided in electronic form by the Fermi LAT team.

\end{acknowledgments}

\bigskip 

\end{document}